\documentclass[11pt,a4paper]{article}
\pdfoutput=1
\usepackage{jcappub}

\usepackage{amsmath,amssymb,amsfonts}
\usepackage{mathtools}
\usepackage{braket}
\usepackage{subfig}
\usepackage{xcolor}
\usepackage{appendix}

\newcommand{\sd}[1]{\mathrm{#1}}

\title{Well-Tempered Cosmology}

\author[a]{William~T.~Emond,}

\author[b]{Chunhao~Li,}

\author[a]{Paul~M.~Saffin,}

\author[c]{Shuang-Yong Zhou}

\affiliation[a]{School of Physics and Astronomy, University of Nottingham,\\
University Park, Nottingham NG7 2RD, United Kingdom}
\affiliation[b]{University of Science and Technology of China, Hefei, Anhui 230026, China}
\affiliation[c]{Interdisciplinary Center for Theoretical Study,\\ University of Science and Technology of China, Hefei, Anhui 230026, China}

\emailAdd{william.emond@nottingham.ac.uk}
\emailAdd{paul.saffin@nottingham.ac.uk}
\emailAdd{lichunha@mail.ustc.edu.cn}
\emailAdd{zhoushy@ustc.edu.cn}

\abstract{
We examine an approach to cosmology, known as Well-Tempering, that allows for a de Sitter phase whose expansion is independent of the cosmological constant. 
Starting from a generic scalar-tensor theory compatible with the recent gravitational wave observation, we impose the Well-Tempering conditions and derive a system that is capable of tuning away the cosmological constant within a sub-class of Horndeski theory, where the scalar has a canonical kinetic term and a general potential. This scenario improves upon the Fab-Four approach by allowing a standard fluid-cosmology before entering the de Sitter phase, and we present an explicit example of our general solution.
}

\begin{document}

\hfill {\footnotesize USTC-ICTS-18-23}

\maketitle


\section{Introduction}
Observational data indicates that our current universe is accelerating, which is one of the outstanding problems beyond the standard cosmological model, and many different approaches have been proposed to attack this problem~\cite{Joyce:2014kja, Copeland:2006wr,Sahni:2006pa}. One of the simplest is to assume that there is a small cosmological constant, i.e., a small vacuum energy, which can fit the current data~\cite{Aghanim:2018eyx}. However, it is difficult to explain a small vacuum energy according to our current understanding of quantum mechanics. The cosmological constant term, as first introduced by Einstein, is allowed in classical general relativity, but quantum field theory typically predicts a huge vacuum energy, which also seems to vary at different stages of the cosmic evolution. Therefore, in the absence of any underlying mechanism, severe and multiple fine-tunings are required to cancel between different contributions of the vacuum energy to produce a viable cosmic history.

Cosmic self-tuning mechanisms have recently been proposed (see, e.g.,~\cite{Charmousis:2011bf, Charmousis:2011ea, Emond:2015efw, Appleby:2011aa, Appleby:2012rx, Linder:2013zoa, Kaloper:2013zca, Kaloper:2014dqa, Kaloper:2014fca, Gubitosi:2011sg, Babichev:2017lmw}), which dynamically remove a large vacuum energy and potentially give rise to the current cosmic acceleration. Weinberg proved a no-go theorem on  various mechanisms designed to remove vacuum energy, which ended earlier attempts on this direction (see~\cite{Weinberg:1988cp} and references therein). The recent self-tuning mechanisms evade the no-go theorem by breaking time translation invariance, which is already spontaneously broken for any interesting cosmic evolution background (see, e.g.,~\cite{Padilla:2015aaa} for a pedagogical introduction of self-tuning and other approaches to the cosmological problem). One could also tune to a de Sitter rather than a Minkowski geometry by, for example, extending the Fab-Four proposal~\cite{Charmousis:2011bf,Martin-Moruno:2015lha}, or trading the fine-tuning of the cosmological constant for a hierarchy of couplings~\cite{Evnin:2018zeo}.  

Scalar fields find many applications in cosmology as they do not carry any spacetime indices and thus naturally admit homogenous and isotropic solutions. Horndeski theory \cite{Horndeski:1974wa} is a general scalar-tensor theory that does not contains higher than second order spacetime derivatives in the equations of motion (thus evading the Ostrogradski instabilities), rediscovered recently as Generalized Galileon theory~\cite{Deffayet:2011gz, Kobayashi:2011nu}, and is the arena where most self-tuning models are implemented.  Recent combined observations on gravitational waves and the accompanying electromagnetic signals have constrained the speed of the gravity to be very close to the speed of light~\cite{TheLIGOScientific:2017qsa, GBM:2017lvd}, and these have been used to eliminate subsets of Horndeski theory as viable gravity models~\cite{Creminelli:2017sry, Sakstein:2017xjx, Ezquiaga:2017ekz, Baker:2017hug, Lombriser:2015sxa, Lombriser:2016yzn, Bettoni:2016mij}, restricting the possible theory space of self-tuning. Note, however, that one should be cautious in one's interpretation of these constraints. As pointed out in Ref.~\cite{Copeland:2018yuh}, it is possible that more general solutions to the equations of motion may evade this constraint. Moreover, it was shown in Ref.~\cite{deRham:2018red}, that the energy scales of the LIGO gravitational wave and electromagnetic signals are very close to the strong coupling scale of these models, viewed as the effective field theories. 

Recently, a ``well-tempered'' self-tuning model has been proposed, which admits a removal of vacuum energies at different epochs of the universe, but leaves the matter sources unaffected, such that it is able to reproduce the standard Big Bang cosmology~\cite{Appleby:2018yci}. This is a virtue of the well-tempered approach that is not shared by previous self-tuning models. Moreover, by construction, it takes into account the constraints placed by the gravitational wave data. However, the scalar field utilized to achieve this goal in the original version has an unusual non-canonical kinetic term. In this paper, we provide a generalization of this model by demanding the presence of a standard quadratic kinetic structure for the scalar field. In doing so, we are able to derive the generic well-tempered form for the Horndeski function $G_{3}$, once we have specified the scalar potential $V(\phi)$, the Horndeski function $G_4(\phi)$, and an arbitrary function of $\phi$ and $X$($=-\frac{1}{2}\nabla^{\mu}\phi\nabla_{\mu}\phi$). We illustrate our general solution with a particular example, and we show that this generalization of the well-tempered model is able to screen the vacuum energy from the space-time curvature, even when subject to phase transitions, as well as support a standard cosmological evolution at early times. 

The layout of this paper is as follows: We start in section \ref{sec:setup} which develops the generic well-tempered solution by imposing the Appleby-Linder degeneracy condition on the subset of Horndeski theory that satisfies the recent gravitational wave astronomy constraints. In order to see how this could work in practise we present a simple example in section \ref{sec:numerical}, showing how the vacuum energy is hidden from the spacetime geometry in various aspects of the cosmic evolution. We end in section \ref{sec:conclusions} with our conclusions.  

\section{Constructing tempered self-tuning models}  
\label{sec:setup}

The arrival of the gravitational wave astronomy has allowed the use of ``multi-messenger'' events to constrain gravity models. In particular, the propagation speed of gravitational degrees of freedom is now constrained to be very close to that of photon. This puts severe constraints on generic scalar-tensor theory. We will work within a subset of Horndeski theory that imposes the phenomenological constraints, and enforce the degeneracy condition of Well-Tempering to obtain the generic Lagrangian that has a canonical kinetic term and a generic potential.

\subsection{Theoretical set-up}
We will start from the full Horndeski Lagrangian:
\begin{equation}\label{eq:horndeski L}
 \mathcal{L}_{H} \ = \ \sum_{i=2}^{5}\:\mathcal{L}_{i}\;,
\end{equation}
where
\begin{subequations}
 \begin{flalign}
  \mathcal{L}_{2} \ =& \ K(\phi,X)\;,\\[0.8em] \mathcal{L}_{3} \ =& \ -G_{3}(\phi,X)\Box\phi\;,\\[0.8em] \mathcal{L}_{4} \ =& \ G_{4}(\phi,X)R +G_{4,X}\Big[(\Box\phi)^{2} \:-\: (\nabla_{\mu}\nabla_{\nu}\phi)^{2}\Big]\;,\\[0.8em] \mathcal{L}_{5} \ =& \ G_{5}(\phi,X)G^{\mu\nu}\nabla_{\mu}\nabla_{\nu}\phi \:-\: \frac{1}{6}G_{5,X}\Big[(\Box\phi)^{3}-3\Box\phi(\nabla_{\mu}\nabla_{\nu}\phi)^{2}+2(\nabla_{\mu}\nabla_{\nu}\phi)^{3}\Big]\;, 
 \end{flalign}
\end{subequations}
and then impose the constraints coming from observations, and the requirements of the well-tempering mechanism. In the above Lagrangian, $R$ and $G^{\mu\nu}$ are the Ricci scalar and Einstein tensor respectively, $K$ and $G_{i}$ ($i = 3,4,5$) are functions of the scalar field $\phi$ and the kinetic term \smash{$X=-\frac{1}{2}\nabla^{\mu}\phi\nabla_{\mu}\phi$}, and we have used the shorthand notation where \smash{$G_{i,\phi}\coloneqq\frac{\partial G_{i}}{\partial\phi}$} and \smash{$G_{i,X}\coloneqq\frac{\partial G_{i}}{\partial X}$}.

In order to guarantee compatibility of the theory with the constraints placed on the Horndeski functions $K$ and $G_{i}$ by the recent multi-messenger gravitational wave data~\cite{Creminelli:2017sry, Sakstein:2017xjx, Ezquiaga:2017ekz, Baker:2017hug, Lombriser:2015sxa, Lombriser:2016yzn, Bettoni:2016mij}, we shall set \mbox{$G_{5}(\phi,X)=0$} and $G_{4}(\phi,X)=G_{4}(\phi)$. As has been mentioned in the introduction, this may in fact be too strong a condition \cite{Copeland:2018yuh, deRham:2018red}.  But, as we shall see, we are able to work within the restriction. With that done, eq.~\eqref{eq:horndeski L} reduces to
\begin{equation}\label{eq:reduced horndeski L}
 \mathcal{L}_{H} \ = \ K(\phi,X) \:-\: G_{3}(\phi,X)\Box\phi \:+\: G_{4}(\phi)R\;.
\end{equation}
As well as the scalar-tensor sector, we also wish to include a description of the matter sector. To do this, we consider a cosmological fluid with corresponding equation of state
\begin{equation}
 w \ = \ \frac{P_{\text{m}}}{\rho_{\text{m}}}
\end{equation}
where $P_{\text{m}}$ and $\rho_{\text{m}}$ are the pressure and density of the fluid, respectively. The full theory is then described by the following action
\begin{equation}\label{eq:full theory}
 S \ = \ \int\:\mathrm{d}^{4}x\:\sqrt{-g}\:\big[\mathcal{L}_{H} \:+\: \mathcal{L}_{\text{m}}\big] 
\end{equation}
where $\mathcal{L}_{\text{m}}$ contains the dynamics of the matter sector.

As we are primarily interested in the cosmological solutions, let us evaluate the theory on an FLRW background:
\begin{equation}
 \mathrm{d}s^{2} \ = \ -N^{2}(t)\mathrm{d}t^{2} \:+\: a^{2}(t)\mathrm{d}\mathbf{x}^{2}\;,
\end{equation}
where $N(t)$ is the lapse function, $a(t)$ is the scale factor, and we adopt the metric signature $(-+++)$ throughout. Taking the scalar field $\phi$ to be homogeneous, i.e., $\phi=\phi(t)$, we find that varying the action [eq.~\eqref{eq:full theory}] with respect to $N(t)$,\footnote{Following the variation with respect to the lapse we subsequently choose the time co-ordinate such that  $N(t)=1$.} and $a(t)$ leads us to introduce
\begin{subequations}
 \begin{flalign}
  \mathcal{E} \ \coloneqq \ \sum_{i=2}^{4}\:\mathcal{E}_{i}\;, \\[0.8em] \mathcal{P} \ \coloneqq \ \sum_{i=2}^{4}\:\mathcal{P}_{i}\;,
 \end{flalign}
\end{subequations}
where
\begin{subequations}\label{eq:E}
 \begin{flalign}
  \mathcal{E}_{2} \ \coloneqq& \ 2XK_{,X} \:-\: K \;, \\[0.8em] \mathcal{E}_{3} \ \coloneqq& \ 6X\sqrt{2X}HG_{3,X} \:-\: 2XG_{3,\phi} \;, \\[0.8em] \mathcal{E}_{4} \ \coloneqq& \ -\: 6H^{2}G_{4} \:-\: 6H\sqrt{2X}G_{4,\phi} \;, 
 \end{flalign}
\end{subequations}
and
\begin{subequations}\label{eq:P}
 \begin{flalign}
  \mathcal{P}_{2} \ \coloneqq& \ K \;, \\[0.8em] \mathcal{P}_{3} \ \coloneqq& \ -\:2X\big(G_{3,\phi} \:+\: \ddot{\phi}G_{3,X}\big) \;, \\[0.8em] \mathcal{P}_{4} \ \coloneqq& \ 2\big(3H^{2} \:+\: 2\dot{H}\big)G_{4} \:+\: 2\big(\ddot{\phi} \:+\: 2H\sqrt{2X}\big)G_{4,\phi} \:+\: 4XG_{4,\phi\phi} \;.
 \end{flalign}
\end{subequations}
We furthermore obtain the continuity equation
\begin{equation}\label{eq:continuity}
 \dot{\rho} \:+\: 3H\big(\rho \:+\: P\big) \ = \ \dot{\rho} \:+\: 3H\rho\big(1 \:+\: w\big) \ = \ 0\;,
\end{equation}
where $H(t)=\frac{\dot{a}(t)}{a(t)}$ is the Hubble parameter. Let us further define the quantities
\begin{subequations}\label{eq:E1 E2}
	\begin{flalign}
	E_{1} \ \coloneqq& \ -\: \mathcal{E} \:-\: \rho \;,\\[0.8em] E_{2} \ \coloneqq& \ \mathcal{P} \:+\: P \;.
	\end{flalign}
\end{subequations}
Then, given the definitions in eqs.~\eqref{eq:E} and~\eqref{eq:P}, the variational equations of motion are 
\begin{subequations}\label{eq:eoms}
	\begin{flalign}
	E_{1} \ =& \ 0\;,\\[0.8em]  E_{2} \ =& \ 0\;.
	\end{flalign}
\end{subequations}
To get the system of equations into appropriate form, we now manipulate eq.~\eqref{eq:eoms} by introducing the following quantities
\begin{subequations}\label{eq:modified eqs.}
 \begin{flalign}
  E^{\phi} \ \coloneqq& \ \dot{E}_{1} \:-\: 3H\big(E_{2} \:-\: E_{1}\big) \;,\\[0.8em] E^{\text{H}} \ \coloneqq& \ E_{2} \:-\: E_{1}\;, \\[0.8em]  E^{\text{F}} \ \coloneqq& \ E_{1}\;.
 \end{flalign}
\end{subequations}
Inserting the expressions in eq.~\eqref{eq:E1 E2} into eq.~\eqref{eq:modified eqs.}, these then have the form
\begin{subequations}\label{eq:scalar and Hubble eom}
 \begin{flalign}
  E^{\phi} \ \coloneqq& \ -\Big[K_{,X} \:+\: 2XK_{,XX} \:-\: 2G_{3,\phi} \:+\: 6H\sqrt{2X}G_{3,X} \:-\: 2XG_{3,\phi X} \:+\: 6HX\sqrt{2X}G_{3,XX}\Big]\ddot{\phi} \nonumber\\[0.2em] &+\: 6\Big[G_{4,\phi} \:-\: XG_{3,X}\Big]\dot{H} \:+\: 6\Big[2G_{4,\phi} \:-\: 3XG_{3,X}\Big]H^{2} \nonumber\\[0.2em] &\:-\: 3\sqrt{2X}\Big[K_{,X} \:-\: 2G_{3,\phi} \:+\: 2XG_{3,\phi X}\Big]H \:+\ 2XG_{3,\phi\phi} \:+\: K_{,\phi} \:-\: 2XK_{,\phi X}  \label{eq:scalar eom} \;,\\[0.8em] 
  E^{\text{H}} \ \coloneqq& \ P \:+\: \rho \:+\: 4G_{4}\dot{H} \:+\: 2\sqrt{2X}\Big[3XG_{3,X} \:-\: G_{4,\phi}\Big]H \:+\: 2\Big[G_{4,\phi} \:-\: XG_{3,X}\Big]\ddot{\phi} \nonumber\\[0.2em] &+\: 2X\Big[K_{,X} \:+\: 2G_{4,\phi\phi} \:-\: 2G_{3,\phi}\Big] \label{eq:Hubble eom}\;, \\[0.8em] 
  E^{\text{F}} \ \coloneqq& \ 6G_{4}H^{2} \:+\: 6\sqrt{2X}\Big[G_{4,\phi} \:-\: XG_{3,X}\Big]H \:+\: 2XG_{3,\phi} \:-\: 2XK_{,X} \:+\: K \:-\: \rho\;.\label{eq:Friedmann constraint}
 \end{flalign}
\end{subequations}
When on-shell these expressions of course satisfy the equations of motion, i.e., $E^{\phi}=0$, $E^{\text{H}}=0$ and $E^{\text{F}}=0$. In the case where $G_{4}=\frac{M^{2}_{\text{Pl}}}{2}$, $G_{3}=0$ and $K=X-V$ we recognise eqs.~\eqref{eq:scalar eom},~\eqref{eq:Hubble eom} and~\eqref{eq:Friedmann constraint} as the standard scalar field and Hubble equations of motion, and Friedmann constraint, respectively (hence the suggestive notation).
\\ 
\subsection{Generic well-tempering}
With the appropriate dynamical equations in place, we are now in a position to construct a tempered self-tuning model, so let us first discuss what we are trying to achieve, and the general problem that self-tuning solutions must overcome. The idea of the self-tuning proposal is that we end up with a geometry,  de Sitter space in our case, that does not depend on the value of the cosmological constant. This necessitates the presence of some sort of degeneracy in the system of equations, as we are trying to get the same geometry, independent of one of the energy sources, namely the vacuum energy. In the Fab-Four proposal \cite{Charmousis:2011bf} this degeneracy was achieved by forcing one of the dynamical equations to be identically solved on the Minkowski solution, which of course is a close relative of de Sitter space. 
While this did indeed tune away the cosmological constant, it was discovered that it did not deal with fluids in a way that would lead to a consistent cosmological history. The approach of Appleby and Linder \cite{Appleby:2018yci} was to instead force two of the dynamical equations to give identical conditions when evaluated on the de Sitter solution. The equations that are chosen to behave in this way are $E^\phi$ and $E^H$ (\ref{eq:scalar eom}, \ref{eq:Hubble eom}), and we have some hope that these will treat the cosmological constant as being special due to the presence of $\rho$ and $P$ in the particular combination $\rho+P$, which vanishes for vacuum energy. Given this, the strategy is to recast eqs.~\eqref{eq:scalar eom} and~\eqref{eq:Hubble eom} as
\begin{subequations}\label{eq: modified scalar and Hubble eom}
 \begin{flalign}
  E^{\phi}_{n}(\phi,\;X,\;H,\;\dot H) \:-\: \ddot{\phi}E^{\phi}_{d}(\phi,\;X,\;H,\;\dot H) \ =& \ 0 \label{eq:modified scalar eom}\;, \;\\[0.8em] 
  E^{\text{H}}_{n}(\phi,\;X,\;H,\;\dot H) \:-\: \ddot{\phi}E^{\text{H}}_{d}(\phi,\;X,\;H,\;\dot H) \ =& \ 0 \label{eq:modified Hubble eom}\;.
 \end{flalign}
\end{subequations}
We then force these two equations to be proportional to one another when we are in a de Sitter phase, namely $H=h=$constant, and $\rho+P=0$, i.e.,
\begin{subequations}\label{eq: de Sitter constraint}
 \begin{flalign}
  E^{\phi}_{n} \ =& \ f(\phi,X)E^{\text{H}}_{n} \;, \;\\[0.8em] E^{\phi}_{d} \ =& \ f(\phi,X)E^{\text{H}}_{d} \;,
 \end{flalign}
\end{subequations}
which leads to 
\begin{equation}\label{eq:numerator/denominator constraint}
 E^{\phi}_{n}E^{\text{H}}_{d} \ = \ E^{\text{H}}_{n}E^{\phi}_{d}\;.
\end{equation}
It is this constraint equation that we now solve. To do so, the only assumption we make at this point is that the scalar field $\phi$ has standard canonical quadratic kinetic structure, and standard potential form:
\begin{equation}
K(\phi,X) \ = \ X \:-\: V(\phi)\;.
\end{equation}
To proceed, we Taylor expand $G_{4}$ and $G_{3}$ and $V(\phi)$ as follows:
\begin{subequations}\label{eq:G4 G3 expansion}
 \begin{flalign}
  G_{4}(\phi) \ =& \ \sum_{n=0}^{\infty}\:c_{n}\,\phi^{n}\;, \\[0.8em] G_{3}(\phi,X) \ =& \ \sum_{m=0}^{\infty}\:f_{m}(X)\,\phi^{m}\;,\label{eq:G3 expansion} \\[0.8em] V(\phi) \ =& \ \sum_{l=0}^{\infty}\:V_{l}\,\phi^{l}\;,
 \end{flalign}
\end{subequations}
where $c_{n}$, and $V_{l}$ are arbitrary constants, and the $f_{m}(X)$ arbitrary functions of \smash{$X=\frac{1}{2}\dot{\phi}^2$}, to be determined. Note that at this point, $G_{4}$ and $G_{3}$ are still completely general analytic functions. Upon equating powers in $\phi$ we are lead to a system of differential equations for the functions $f_{m}(X)$. In practice, it is not possible to solve this infinite number of equations and, as such, we truncate the sums in eq.~\eqref{eq:G4 G3 expansion} to finite order. In doing so, it is possible to determine explicit analytic expressions for a finite number of coefficient functions $f_{m}(X)$ (see appendix~\ref{sec:G3 calc} for a detailed example calculation). One can then resum these $f_{m}(X)$ to all orders, such that we obtain an expression for $G_{3}$ that is no longer in the form of a Taylor expansion. We find,
\begin{flalign}\label{eq:G3 sol}
 G_{3}(\phi,X) \ =& \ -\:\frac{X \:-\:V(\phi)}{3h\sqrt{2X}} \:-\: \frac{2h\:G_{4}(\phi)}{\sqrt{2X}} \:+\: \frac{2}{3}\:\frac{\partial}{\partial\phi}\mathcal{G}(u)\text{log}\bigg(\frac{X}{\mu^{4}}\bigg) \nonumber\\[0.4em] &\:-\: \frac{4}{3}\:\int_{0}^{1}\:\frac{\frac{\partial}{\partial\phi}\mathcal{G}(u) \:-\: \frac{\partial}{\partial\phi}\mathcal{G}(\tilde{u})}{1-s}\:\mathrm{d}s \: +\: F(\phi,X)   \;,
\end{flalign}
where $\mu$ is an arbitrary mass scale introduced to make the log term dimensionally correct, and we have introduced the variables $u$ and $\tilde{u}$, defined respectively as
\begin{equation}\label{eq:variables def}
u \ \coloneqq \ \phi \:+\: \frac{\sqrt{2X}}{3h}\;, \qquad \tilde{u} \ \coloneqq \ \phi \:+\: s\frac{\sqrt{2X}}{3h}\;.
\end{equation}
The functions $\mathcal{G}$ and $F$ are defined as
\begin{subequations}\label{eq:function defs}
 \begin{flalign}
  \mathcal{G}(\phi) \ \coloneqq& \ G_{4}(\phi) \:+\: \frac{1}{12h^{2}}V(\phi)\;, \\[0.8em] F(\phi,X) \ \coloneqq& \ \frac{a_{0}h}{\sqrt{2X}} \:+\: \mathcal{F}(u)\;,
 \end{flalign}
\end{subequations}
with $a_{0}$ an arbitrary constant, and $\mathcal{F}$ an arbitrary function. This final expression for $G_{3}$ [eq.~\eqref{eq:G3 sol}] is fully consistent with the well-tempered constraint [eq.~\eqref{eq:numerator/denominator constraint}], as can be shown by inserting it directly into eq.~\eqref{eq:numerator/denominator constraint}. This is the main result of the paper.  

This final form of $G_{3}$ [eq.~\eqref{eq:G3 sol}] is a pleasing result, as it is expressed in terms of arbitrary functions $G_{4}(\phi)$, $V(\phi)$ and $F(\phi,X)$, which leaves a large room for potential applications. The task at hand now, is to determine whether there exist suitable choices of $G_{4}(\phi)$, $V(\phi)$ and $F(\phi,X)$, such that the model admits self-tuning solutions. With this in mind, we shall now study a particular example model. 
\\
\section{An example}
\label{sec:numerical}
Let us start by considering the following example model:
\begin{subequations}\label{eq:working exp. choices}
 \begin{flalign}
  G_{4}(\phi) \ =& \ \frac{M_{\text{Pl}}^{2}}{2} \:-\: \frac{1}{12h^{2}}V(\phi)\;,\label{eq:G4 choice} \\[0.8em] \mathcal{F}(u) \ =& \ 0 \;, \\[0.8em] a_{0} \ =& \ M_{\text{Pl}}^{2}\;,
 \end{flalign}
\end{subequations}
where we keep the scalar potential $V(\phi)$ arbitrary for the time being. Referring back to eq.~\eqref{eq:G3 sol}, one finds that in this case $G_{3}$ takes the remarkably simple form
\begin{equation}\label{eq:G3 quad pot example}
 G_{3}(\phi,X) \ = \ \frac{1}{6h\sqrt{2X}}\Big[3V(\phi) \:-\: 2X \Big]\;.
\end{equation}
Given that we wish to determine the behaviour of the system as the Hubble parameter tends towards its late-time de Sitter solution $H=h$, it is instructive to expand $H$ around this point such that $H(t)=h+\delta H(t)$, where $\delta H(t)$ is a time-dependent perturbation around the de Sitter point. Given this, the expressions $E^{\text{F}}$, $E^{\text{H}}$ and $E^{\phi}$ (cf.~eq.~\eqref{eq:scalar and Hubble eom}) can be recast as follows:
\begin{subequations}\label{eq:eoms general V}
 \begin{flalign}
  E^{\text{F}} \ \coloneqq& \ \bigg\lbrace 3h^{2}M_{\text{Pl}}^{2} \:-\: \frac{V(\phi)}{2} \bigg\rbrace\bigg(\frac{\delta H}{h}\bigg)^{2} \:+\: \bigg\lbrace 6h^{2}M_{\text{Pl}}^{2} \:+\: \frac{V(\phi)}{2} \:-\: \frac{\dot{\phi}V'(\phi)}{2h} \:+\: \frac{\dot{\phi}^{2}}{2} \bigg\rbrace\frac{\delta H}{h} \nonumber\\[0.2em] &\ +\: 3M_{\text{Pl}}^{2}h^{2} \:-\: \rho\;,\label{eq:Friedmann constraint general V} \\[0.8em] E^{\text{H}} \ \coloneqq& \ P \:+\: \rho \:+\: \bigg\lbrace 2hM_{\text{Pl}}^{2} \:-\: \frac{V(\phi)}{3h} \bigg\rbrace\frac{\mathrm{d}}{\mathrm{d}t}\bigg(\frac{\delta H}{h}\bigg) \:+\: \bigg\lbrace \frac{\dot{\phi}V'(\phi)}{6h} \:-\: \frac{3V(\phi)}{2} \:-\: \frac{\dot{\phi}^{2}}{2}\bigg\rbrace\frac{\delta H}{h} \nonumber\\[0.2em] &\ +\: \bigg\lbrace \frac{\dot{\phi}}{6h} \:+\: \frac{V(\phi)}{2h\dot{\phi}} \:-\: \frac{V'(\phi)}{6h^{2}} \bigg\rbrace\ddot{\phi} \:+\: \frac{1}{2}\bigg\lbrace 1 \:-\: \frac{V''(\phi)}{3h^{2}} \bigg\rbrace\dot{\phi}^{2} \:-\: \frac{5\dot{\phi}V'(\phi)}{6h} \:-\: \frac{3V(\phi)}{2} \;,\label{eq:Hubble eom general V} \\[0.8em] E^{\phi} \ \coloneqq& \ -\: \frac{1}{2}\bigg\lbrace 1 \:+\: \frac{3V(\phi)}{\dot{\phi}^{2}} \:-\: \frac{V'(\phi)}{h\dot{\phi}} \bigg\rbrace\ddot{\phi} \:+\: \frac{1}{2}\bigg\lbrace \frac{V''(\phi)}{h} \:-\: 3h \bigg\rbrace\dot{\phi} \:+\: \frac{9hV(\phi)}{2\dot{\phi}} \:+\: \frac{5V'(\phi)}{2} \nonumber\\[0.2em] &\ +\: \bigg\lbrace \frac{\dot{\phi}}{2} \:+\: \frac{3V(\phi)}{2\dot{\phi}} \:-\: \frac{V'(\phi)}{2h}\bigg\rbrace\frac{\mathrm{d}}{\mathrm{d}t}\bigg(\frac{\delta H}{h}\bigg) \:+\: \bigg\lbrace \frac{3h\dot{\phi}}{2} \:+\: \frac{9hV(\phi)}{2\dot{\phi}} \:-\: V'(\phi) \bigg\rbrace\bigg(\frac{\delta H}{h}\bigg)^{2}\nonumber\\[0.2em] &\ +\: \bigg\lbrace \bigg[\frac{1}{2} \:-\: \frac{3V(\phi)}{2\dot{\phi}^{2}}\bigg]\ddot{\phi} \:+\: \frac{9hV(\phi)}{\dot{\phi}} \:+\: \frac{5V'(\phi)}{2} \bigg\rbrace\frac{\delta H}{h}\;.\label{eq:scalar eom general V} 
 \end{flalign}
\end{subequations}
So as not to overcomplicate the discussion, we shall now consider the case in which the scalar potential is quadratic in form, i.e.
\begin{equation}
 V(\phi) \ = \ \frac{m^{2}}{2}\phi^{2}\;.
\end{equation}
Given this choice for $V$, eqs.~\eqref{eq:Friedmann constraint general V},~\eqref{eq:Hubble eom general V} and~\eqref{eq:scalar eom general V} take the form 
\begin{subequations}\label{eq:quad pot example}
 \begin{flalign}
  E^{\text{F}} \ \coloneqq& \ \bigg\lbrace 3h^{2}M_{\text{Pl}}^{2} \:-\: \frac{m^{2}\phi^{2}}{4} \bigg\rbrace\bigg(\frac{\delta H}{h}\bigg)^{2} \:+\: \bigg\lbrace 6h^{2}M_{\text{Pl}}^{2} \:+\: \frac{m^{2}\phi^{2}}{4} \:-\: \frac{m^{2}\phi\dot{\phi}}{2h} \:+\: \frac{\dot{\phi}^{2}}{2} \bigg\rbrace\frac{\delta H}{h} \nonumber\\[0.2em] &\ +\: 3M_{\text{Pl}}^{2}h^{2} \:-\: \rho\;, \label{eq:Friedmann constraint quad pot}\\[0.8em] E^{\text{H}} \ \coloneqq& \ P \:+\: \rho \:+\: \bigg\lbrace 2hM_{\text{Pl}}^{2} \:-\: \frac{m^{2}\phi^{2}}{6h} \bigg\rbrace\frac{\mathrm{d}}{\mathrm{d}t}\bigg(\frac{\delta H}{h}\bigg) \:+\: \bigg\lbrace \frac{m^{2}\phi\dot{\phi}}{6h} \:-\: \frac{3m^{2}\phi^{2}}{4} \:-\: \frac{\dot{\phi}^{2}}{2} \bigg\rbrace\frac{\delta H}{h} \nonumber\\[0.2em] &\ +\: \bigg\lbrace \frac{\dot{\phi}}{6h} \:+\: \frac{m^{2}\phi^{2}}{4h\dot{\phi}} \:-\: \frac{m^{2}\phi}{6h^{2}} \bigg\rbrace\ddot{\phi} \:+\: \frac{1}{2}\bigg\lbrace 1 \:-\: \frac{m^{2}}{3h^{2}} \bigg\rbrace\dot{\phi}^{2} \:-\: \frac{5m^{2}\phi\dot{\phi}}{6h} \:-\: \frac{3m^{2}\phi^{2}}{4} \;,\label{eq:Hubble eom quad pot} \\[0.8em] E^{\phi} \ =& \ -\: \frac{1}{2}\bigg\lbrace 1 \:+\: \frac{3m^{2}\phi^{2}}{2\dot{\phi}^{2}} \:-\: \frac{m^{2}\phi}{h\dot{\phi}} \bigg\rbrace\ddot{\phi} \:+\: \frac{1}{2}\bigg\lbrace \frac{m^{2}}{h} \:-\: 3h \bigg\rbrace\dot{\phi} \:+\: \frac{9hm^{2}\phi^{2}}{4\dot{\phi}} \:+\: \frac{5m^{2}\phi}{2} \nonumber\\[0.2em] &\ +\: \bigg\lbrace \frac{\dot{\phi}}{2} \:+\: \frac{3m^{2}\phi^{2}}{4\dot{\phi}} \:-\: \frac{m^{2}\phi}{2h}\bigg\rbrace\frac{\mathrm{d}}{\mathrm{d}t}\bigg(\frac{\delta H}{h}\bigg) \:+\: \bigg\lbrace \frac{3h\dot{\phi}}{2} \:+\: \frac{9hm^{2}\phi^{2}}{4\dot{\phi}} \:-\: m^{2}\phi \bigg\rbrace\bigg(\frac{\delta H}{h}\bigg)^{2}\nonumber\\[0.2em] &\ +\: \bigg\lbrace \bigg[\frac{1}{2} \:-\: \frac{3m^{2}\phi^{2}}{4\dot{\phi}^{2}}\bigg]\ddot{\phi} \:+\: \frac{9hm^{2}\phi^{2}}{2\dot{\phi}} \:+\: \frac{5m^{2}\phi}{2} \bigg\rbrace\frac{\delta H}{h}\;.\label{eq:scalar eom quad pot} 
 \end{flalign}
\end{subequations}
\\
\subsection{Vacuum energy screening}
We now proceed to consider the vacuum limit, in which the energy density $\rho$ is dominated by a (constant) vacuum energy density $\rho_{\Lambda}$ - the cosmological constant. This amounts to setting $P=-\rho$, with $\rho= \rho_{\Lambda}$ in eqs.~\eqref{eq:Friedmann constraint quad pot} and~\eqref{eq:Hubble eom quad pot}. To solve the system of equations [eq.~\eqref{eq:quad pot example}] numerically, we first solve the Friedmann constraint [eq.~\eqref{eq:Friedmann constraint quad pot}] and Hubble equation of motion [eq.~\eqref{eq:Hubble eom quad pot}] for $\delta H$ and ${\rm d}(\delta H)/{\rm d} t$, respectively. We then insert these expressions back into the scalar equation of motion [eq.~\eqref{eq:scalar eom quad pot}]. The advantage of adopting this approach is that we then only have to solve one differential equation of motion, which itself automatically accounts for the Friedmann constraint. 

To proceed, we choose the free parameters of the model as follows:
\begin{equation}\label{eq:parameter choices}
 h \ = \ 10^{-3}M_{\text{Pl}} \;;\qquad m \ = \ 10^{-4}M_{\text{Pl}}\;;\qquad \rho_{\Lambda} \ = \ M_{\Lambda}^{4} \ = \ 10^{-4}M_{\text{Pl}}^{4}\;.
\end{equation}
Note that the logarithmic contributions to $G_{3}$ (cf.~eq.~\eqref{eq:G3 quad pot example}) drop out in this case and thus we do not need to specify a value for the mass scale $\mu$. This choice presents a representative example in which there is a modest hierarchy between the mass scales present in the model $M_{\text{Pl}}>M_{\Lambda}(>m)$. Of course, in reality one expects this hierarchy to be much larger if $M_{\Lambda}$ sets the mass scale associated with the late time acceleration of the universe. 
\begin{figure}[t!]
	\begin{center}
		\hfill\hfill\subfloat[]{\includegraphics[scale=0.75]{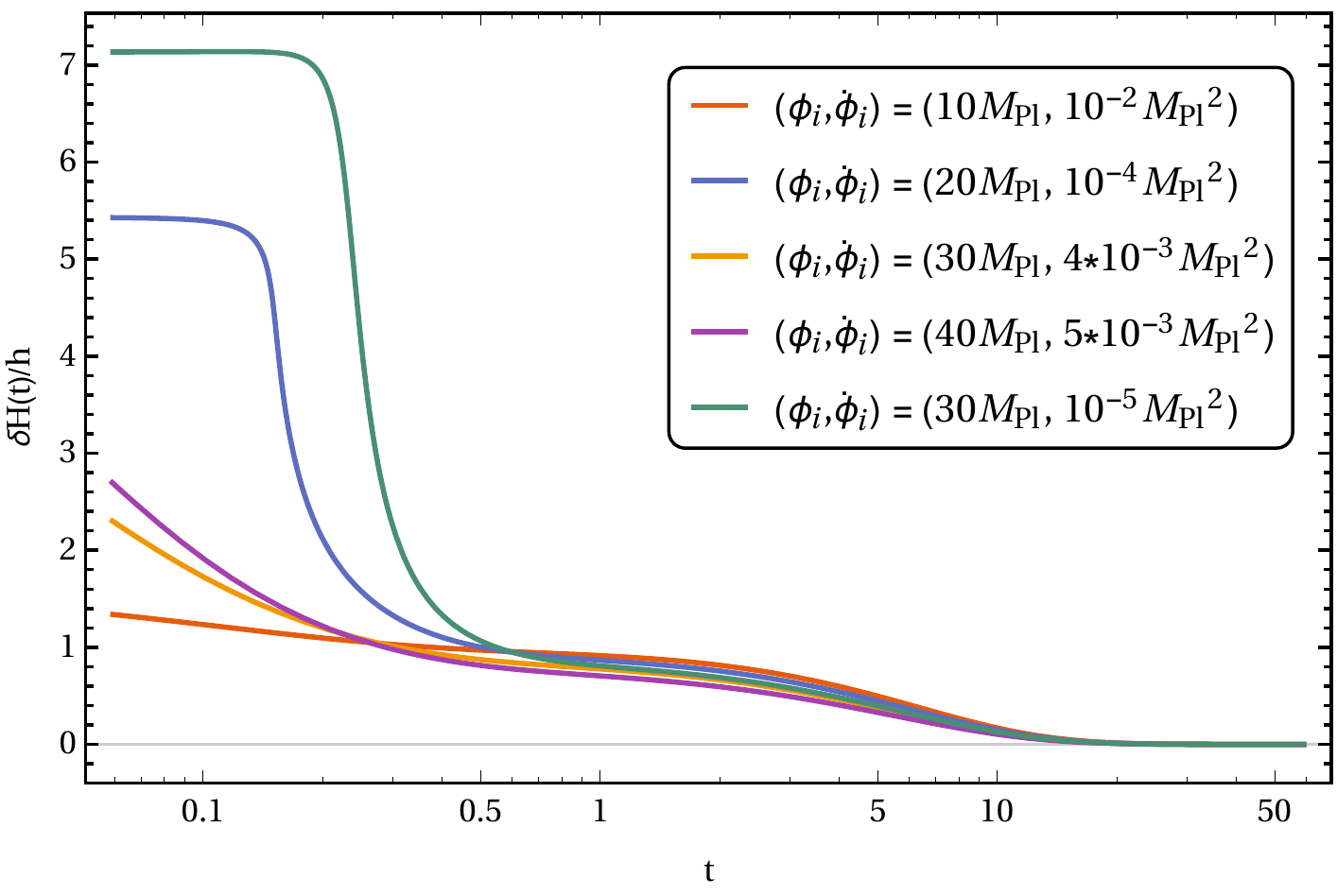} \label{fig:quad pot exp. deltaH evo}}\hfill\hfill\vspace{2em} \subfloat[]{\includegraphics[scale=0.815]{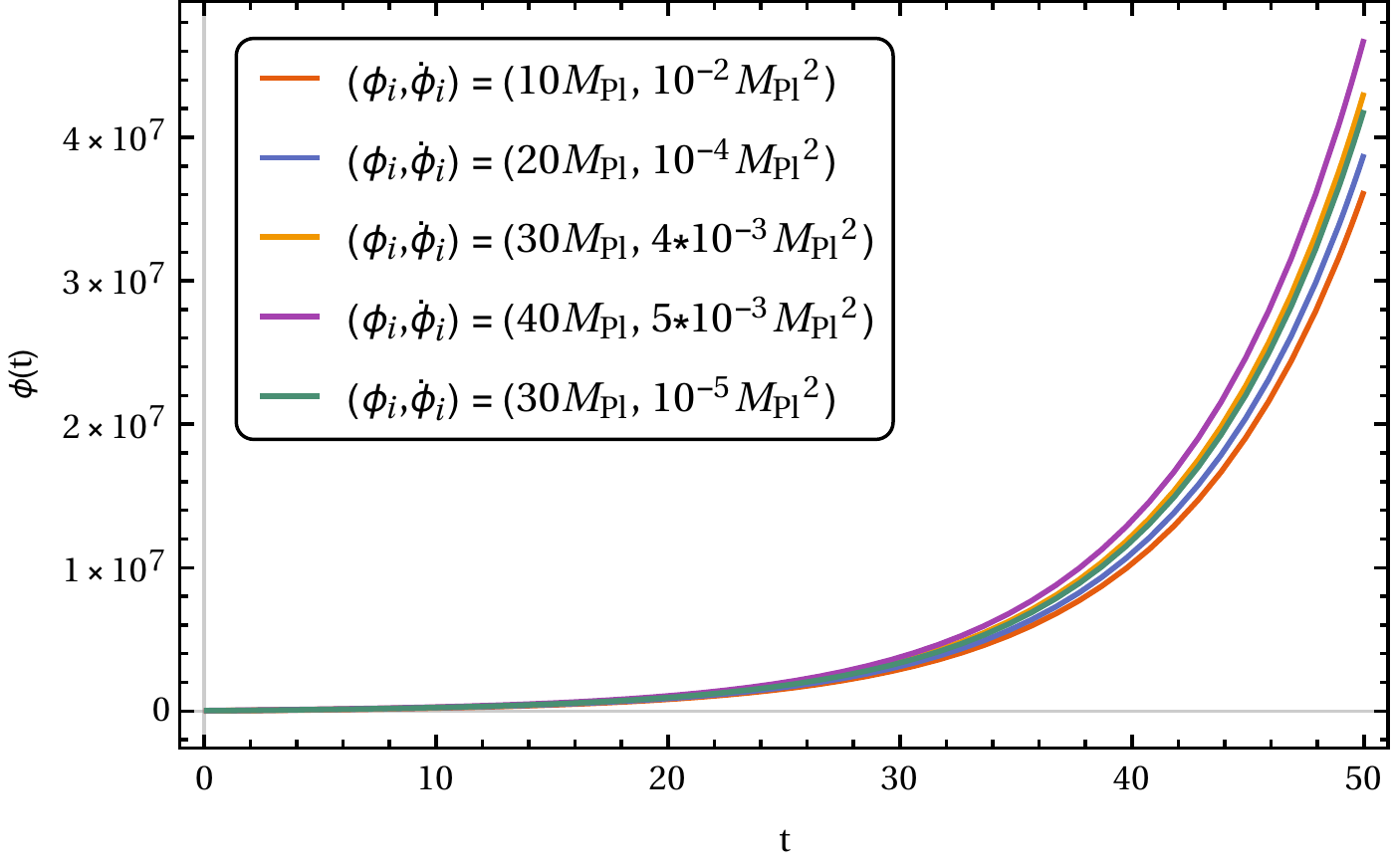} \qquad\label{fig:quad pot exp. phi evo}}
	\end{center}
	\caption{(a) Plot of the evolution of $\frac{\delta H}{h}$, and (b) evolution of $\phi$ for various choices of initial conditions for $\phi$ and $\dot{\phi}$ (where $t$ is given in units of $1/M$, where $M=10^{-3}M_{\text{Pl}}$).}\label{fig:quad pot exp.}
\end{figure}

We are now in a position to solve the system of equations [eq.~\eqref{eq:quad pot example}] numerically. We do so for several different choices of initial data $(\phi(t_{i}),\,\dot{\phi}(t_{i}))=(\phi_{i},\,\dot{\phi}_{i})$. Note that, as we have incorporated the Friedmann constraint into the equation of motion we are now free to choose $\phi$ and $\dot{\phi}$ essentially arbitrarily (in practice, this is limited by whether the system is numerically stable for a given choice) and this fixes the initial value of $H$. Our choices for the initial data $(\phi_{i},\,\dot{\phi}_{i})$ are as follows: \\ \\ \\
\begin{subequations}\label{eq:initial conditions}
 \begin{flalign}
  (\phi_{i},\dot{\phi}_{i})\ =& \ (10M_{\text{Pl}},\,10^{-2}M_{\text{Pl}}^{2})\;, \\[0.2em] (\phi_{i},\dot{\phi}_{i})\ =& \ (20M_{\text{Pl}},\,10^{-4}M_{\text{Pl}}^{2})\;, \\[0.2em] (\phi_{i},\dot{\phi}_{i})\ =& \ (30M_{\text{Pl}},\, 4\times 10^{-3}M_{\text{Pl}}^{2})\;, \\[0.2em] (\phi_{i},\dot{\phi}_{i})\ =& \ (40M_{\text{Pl}},\,5\times 10^{-3}M_{\text{Pl}}^{2})\;, \\[0.2em] (\phi_{i},\dot{\phi}_{i})\ =& \ (30M_{\text{Pl}},\,10^{-5}M_{\text{Pl}}^{2})\;. 
 \end{flalign}
\end{subequations}
In Fig.~\ref{fig:quad pot exp. deltaH evo} are the plots of the evolution of $\frac{\delta H}{h}$ for the various choices of initial data for $\phi$ and $\dot{\phi}$ given in eq.~\eqref{eq:initial conditions}. We see in each case that $\frac{\delta H}{h}\to 0$, i.e. $H\to h$ in the late-time limit. The fact that $H$ asymptotes to $h$, but never reaches exactly $H=h$ is due to an important, but subtle, point that we shall elaborate on later. Figure~\ref{fig:quad pot exp. phi evo} shows the corresponding evolution of $\phi$ for each choice of initial data in this example. In particular, we see that the scalar field undergoes an exponential growth at late times.  

\begin{figure}[t!]
 \begin{center}
   \includegraphics[scale=0.85]{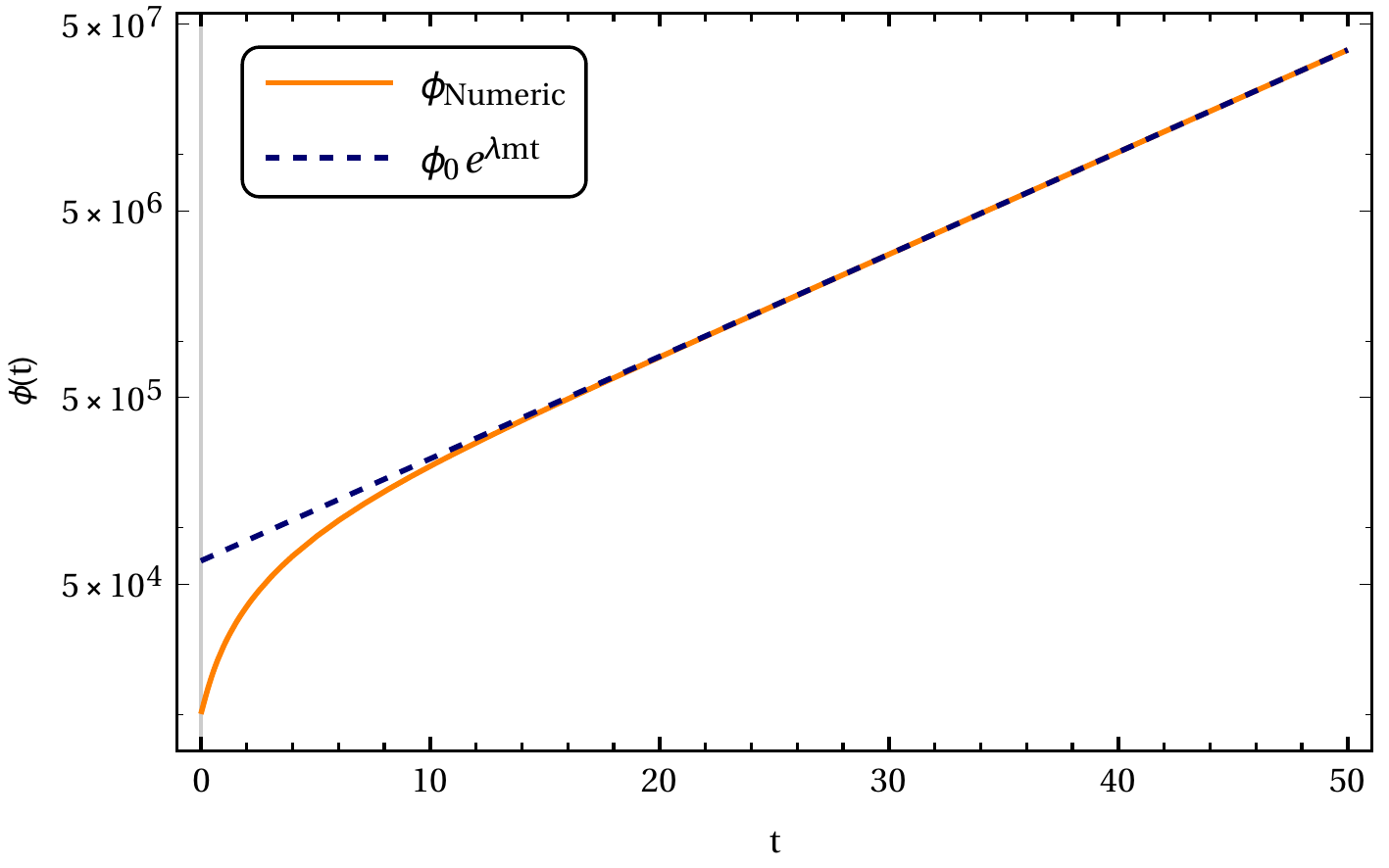} 
  \end{center}
  \caption{Comparison of the characteristic numerical solution for $\phi$ (solid orange line), determined from the full equations of motion, and the corresponding approximate analytic solution (dashed blue line) $\phi_{0}e^{\lambda mt}$, with $\phi_{0}\approx 6.61\times 10M_{\text{Pl}}$ in this case. We see that they match up very well at late times.}\label{fig:numeric analytic scalar}
\end{figure}

Given that the data suggests that the scalar field $\phi$ is able to self-tune, we now wish to determine an approximate analytic solution for its asymptotic, late-time behaviour, to check that $H=h$ is indeed an attractor solution. To do so, we first appeal to the numerical results. These suggest that at late times we are safe to drop all of the terms depending on $\delta H$ and $\frac{\mathrm{d}}{\mathrm{d} t}(\delta H)$ appearing in the scalar equation of motion [eq.~\eqref{eq:scalar eom quad pot}]. These terms decay rapidly over time. Referring back to eq.~\eqref{eq:scalar eom quad pot}, we see then that the approximate equation of motion for $\phi$ is:
\begin{equation}\label{eq:approx scalar eom}
 E^{\phi} \ \approx \ -\: \frac{1}{2}\bigg\lbrace 1 \:+\: \frac{3m^{2}\phi^{2}}{2\dot{\phi}^{2}} \:-\: \frac{m^{2}\phi}{h\dot{\phi}} \bigg\rbrace\ddot{\phi} \:+\: \frac{1}{2}\bigg\lbrace \frac{m^{2}}{h} \:-\: 3h \bigg\rbrace\dot{\phi} \:+\: \frac{9hm^{2}\phi^{2}}{4\dot{\phi}} \:+\: \frac{5m^{2}\phi}{2} \;.
\end{equation}
It is clear that when on-shell, eq.~\eqref{eq:approx scalar eom} admits exponential solutions. Let us therefore try the ansatz
\begin{equation}
 \phi(t) \ \approx \ \phi_{0}\:e^{\alpha t}\;,
\end{equation}
where $\phi_{0}$ and $\alpha$ are constants to be determined. Inserting this ansatz into eq.~\eqref{eq:approx scalar eom} leads us to the following cubic equation
\begin{equation}\label{eq:alpha eq.}
 \alpha^{3} \:+\: \Big(3h \:-\: \frac{2m^{2}}{h}\Big)\alpha^{2} \:-\: \frac{7m^{2}}{2}\alpha \:-\: \frac{9hm^{2}}{2} \ \approx \ 0\;.
\end{equation}
The solutions to this equation are not particularly elegant in form, but do admit one positive solution. We shall discard the two remaining negative solutions as we are only concerned with the asymptotic behaviour of $\phi$ and these will clearly rapidly decay. As such, we are left with the following approximate solution for $\alpha$:
\begin{equation}
 \alpha \ \approx \ \lambda m\;,
\end{equation}
where $\lambda\approx 1.2613$. Given this, the approximate late-time behaviour of $\phi$ is of the form
\begin{equation}\label{eq:approx phi}
 \phi(t) \ \approx \ \phi_{0}\:e^{\lambda mt}\;.
\end{equation}
One can show that, for a suitable choice of $\phi_{0}$, this approximate analytic solution for $\phi$ agrees very well with the asymptotic behviour of the numerical solution for each choice of initial data. See Fig.~\ref{fig:numeric analytic scalar}, for example.

Let us now return to eq.~\eqref{eq:Friedmann constraint quad pot} and solve the on-shell Friedmann constraint in terms of $\delta H$:
\begin{flalign}\label{eq:deltaH sol}
 \delta H(t) \ =& \ \mathcal{A}^{-1}(\phi)\mathcal{B}(\phi,X)  \:-\: \mathcal{A}^{-1}(\phi)\bigg[\mathcal{B}^{2}(\phi,X) \:+\: 2\mathcal{A}(\phi)\Big[3h^{2}M_{\text{Pl}}^{2} \:-\: \rho_{\Lambda}\Big]\bigg]^{1/2}\;,
\end{flalign}
where
\begin{subequations}
 \begin{flalign}
  \mathcal{A}(\phi) \ =& \ \frac{m^{2}\phi^{2}}{2h^{2}} \:-\: 6M_{\text{Pl}}^{2} \;, \\[0.8em]
  \mathcal{B}(\phi,X) \ =& \ 6hM_{\text{Pl}}^{2} \:+\: \frac{X}{h} \:+\: \frac{m^{2}\phi^{2}}{4h} \:-\: \frac{m^{2}\phi\sqrt{2X}}{2h^{2}}\;.
 \end{flalign}
\end{subequations}
Inserting the late-time analytic solution for $\phi$ [eq.~\eqref{eq:approx phi}] into eq.~\eqref{eq:deltaH sol}, it takes the following form:
\begin{flalign}\label{eq:deltaH sol2}
 \delta H(t) \ & \to \ \frac{\phi_{0}^{2}}{\frac{m^{2}\phi_{0}^{2}}{2h^{2}} \:-\: 6M_{\text{Pl}}^{2}\:e^{-2\lambda mt}}\Bigg\lbrace a \:+\: \frac{6hM_{\text{Pl}}^{2}}{\phi_{0}^{2}}\:e^{-2\lambda mt} \nonumber\\[0.2em] & -\: \sqrt{\Big[a \:+\: \frac{6hM_{\text{Pl}}^{2}}{\phi_{0}^{2}}\:e^{-2\lambda mt}\Big]^{2} \:-\: \frac{e^{-2\lambda mt}}{\phi_{0}^{2}}\Big[\frac{m^{2}}{h^{2}} \:-\: \frac{12M_{\text{Pl}}^{2}}{\phi_{0}^{2}}\:e^{-2\lambda mt}\Big]\Big[3h^{2}M_{\text{Pl}}^{2} \:-\: \rho_{\Lambda}\Big]}\Bigg\rbrace\;.
\end{flalign}
where
\begin{subequations}
 \begin{flalign}
  a \ =& \ \frac{\lambda^{2} m^{2}}{2h^{2}}\Big[1 \:+\: \frac{h}{2\lambda^{2}} \:-\: \frac{m}{\lambda}\Big]\;.
 \end{flalign}
\end{subequations}
We see then, that $\delta H$ undergoes an exponential decay, with only residual finite pieces remaining in the late-time limit. In fact, these residual contributions conspire to cancel exactly to zero. Indeed, in the late-time limit of this expression, i.e., as $t\to\infty$, one can neglect the exponentially decaying contributions, such that one is left with: 
\begin{equation}\label{eq:deltaH late time limit}
 \delta H(t) \ \to \ \frac{2h^{2}}{m^{2}}\Bigg\lbrace \frac{\lambda^{2} m^{2}}{2h^{2}}\Big[1 \:+\: \frac{h}{2\lambda^{2}} \:-\: \frac{m}{\lambda}\Big] \:-\: \frac{\lambda^{2} m^{2}}{2h^{2}}\Big[1 \:+\: \frac{h}{2\lambda^{2}} \:-\: \frac{m}{\lambda}\Big]\Bigg\rbrace \ = \ 0\;.
\end{equation}
We see therefore, that $H=h$ is an attractor solution and thus this model admits well-tempered solutions, at least in the vacuum case, where the cosmological constant dominates the energy density. Moreover, it is evident from eq.~\eqref{eq:deltaH sol2}, that the contributions to $\delta H$ from $\rho_{\Lambda}$ are exponentially suppressed relative to the residual constant pieces, such that $\rho_{\Lambda}$ very quickly becomes irrelevant in the late-time evolution of $\delta H$. An important consequence of this, is that one does not need to re-adjust the value of $\phi_{0}$ due to changes in $\rho_{\Lambda}$. Once one has fixed $\phi_{0}$ for a given value of $\rho_{\Lambda}$, then $\delta H$ will always exhibit the late-time behaviour given by eqs.~\eqref{eq:deltaH sol2} and~\eqref{eq:deltaH late time limit}. Essentially, this is due to the fact that any change in $\phi_{0}$ is equivalent to shifting the initial time $t_{i}$, which of course, the late-time solutions are insensitive to. This will become clear when we sudy the effects of a phase transition in the vacuum energy.

Notice that in our numerical analysis in this subsection, we never started out the Hubble parameter on the de Sitter solution $H=h$. In fact, although $H=h$ is an attractor solution in the late-time limit, for this example model, is not possible for the Hubble parameter to start out at the de Sitter solution $H=h$, but only ever asymptote to it. Indeed, for any model in which $G_{4}$ is given by eq.~\eqref{eq:G4 choice}, or \smash{$G_{4}={M_{\text{Pl}}^{2}}/{2}$} and $V(\phi)$ is an arbitrary potential, this will be the case. The reason for this is somewhat subtle and has to do with the competing effects between the growth of the scalar field $\phi$ (and its time-derivatives $\dot{\phi}$ and $\ddot{\phi}$), and the decay of $\delta H$. To see why this is the case, let us return to the Friedmann constraint [eq.~\eqref{eq:Friedmann constraint general V}] for a general potential $V(\phi)$, and set \smash{$\delta H = \frac{\mathrm{d}}{\mathrm{d}t}\big(\delta H\big)=0$}. This is equivalent to setting $H=h$, and in doing so, eq.~\eqref{eq:Friedmann constraint general V} becomes  
\begin{equation}
 E^{\text{F}} \ = \ 3M_{\text{Pl}}^{2}h^{2} \:-\: \rho_{\Lambda}\;.
\end{equation}
It appears that the scalar field no longer impacts on the dynamics of the Hubble parameter, leaving us no better off than we were before, as it seems that we are forced into accepting \smash{$h^2=\frac{1}{3M_{\text{Pl}}^2}\rho_\Lambda$}. However, we have shown that it \emph{is} possible to obtain tempered self-tuning solutions for this class of models. The solution to this problem lies in the fact that we cannot naively set $H=h$ and then disregard terms proportional to $(H-h)$. Indeed, such terms are typically of the form $(H-h)\beta(\phi,X)$, where $\beta(\phi,X)$ is some function of $\phi$ and $X$. In the limit $H\to h$, $\beta(\phi,X)$ will generically increase, whilst $(H-h)\to 0$. This results in residual \emph{finite} contributions to the Friedmann constraint that can partially cancel with $\rho_{\Lambda}$. One can conduct the same analysis for the case where \smash{$G_{4}=\frac{M_{\text{Pl}}^{2}}{2}$} and arrive at the same conclusions.  
\\
\subsection{Including matter}
In the previous subsection we have been able to show numerically that for the example model eq.~\eqref{eq:eoms general V}, tempered self-tuning solutions exist for a quadratic scalar potential in the vacuum case, where the cosmological constant dominates the energy density, i.e., $P=-\rho_{\Lambda}$. Furthermore, we found an approximate analytic solution for the late-time behaviour of the scalar field, from which we were able to show that $H=h$ is an attractor solution. However, we would also like to allow for a viable cosmological history away from the de Sitter state, involving suitably long periods of radiation and matter domination. For this to be possible, we require that the tempered self-tuning solutions screen only the vacuum energy from the spacetime curvature, and not all forms of matter and radiation present. 

To analyse this scenario, we return to the original equations for this example model [eq.~\eqref{eq:quad pot example}], including a generic matter component, such that
\begin{subequations}
 \begin{flalign}
  \rho \ =&\ \rho_{\text{mat}} \:+\: \rho_{\Lambda}\;,\label{eq:density matter}\\[0.8em]
  P \ =& \ w_\text{mat}\rho_{\text{mat}} \:-\: \rho_{\Lambda}\;,\label{eq:pressure matter}
 \end{flalign}
\end{subequations}
\begin{figure}[t!]
 \begin{center}
   \includegraphics[scale=0.87]{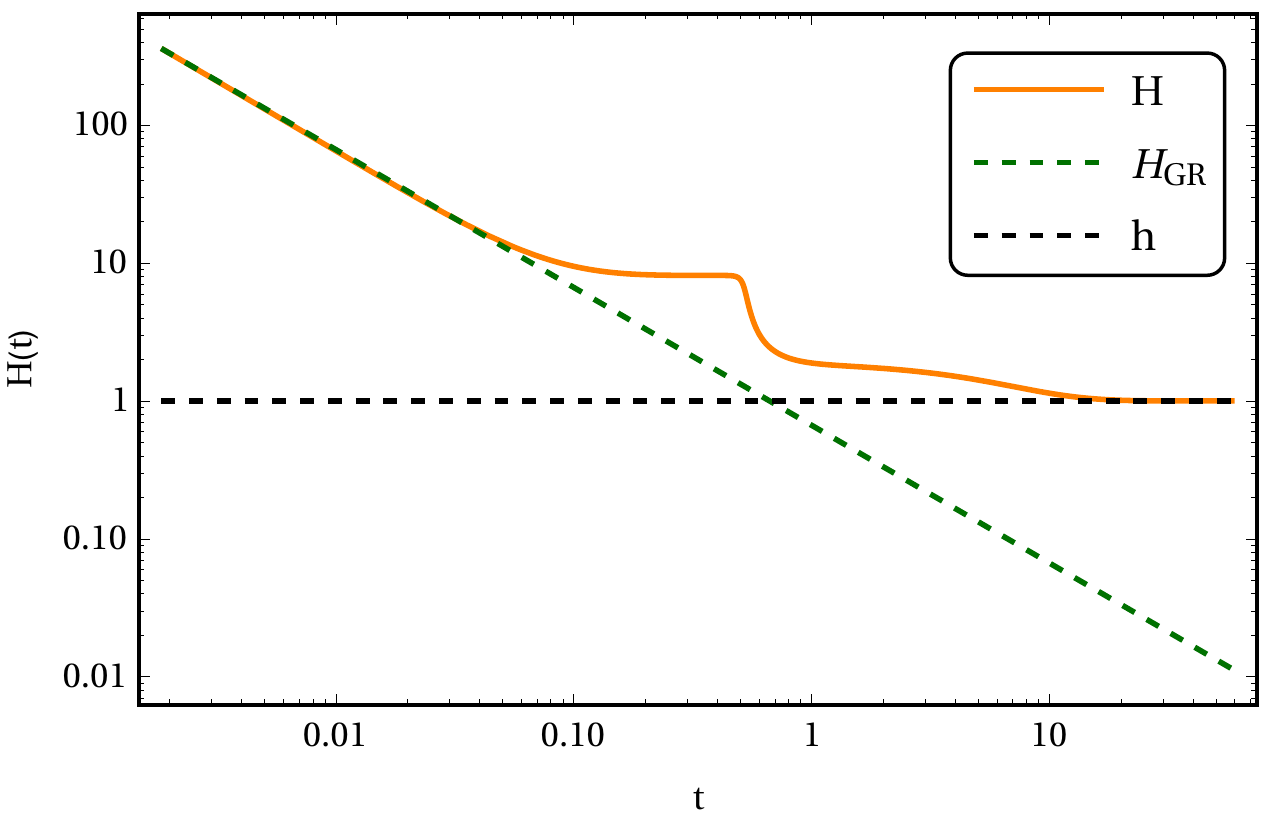} 
  \end{center}
  \caption{Here, we show the evolution of $H$ (solid orange line), numerically evolved using eqs.~\eqref{eq:quad pot example} and~\eqref{eq:continuity} (where $\rho$ and $P$ are given by eqs.~\eqref{eq:density matter} and~\eqref{eq:pressure matter}, respectively). The dashed green line indicates the GR solution, i.e., $H_{\text{GR}}=\frac{2}{3t}$, and the dashed black line is the de Sitter point $H=h$. We see that at early times $H$ exhibits a matter dominated period, for which $H\simeq H_{\text{GR}}$, followed by an approach towards the de Sitter point $H=h$.} \label{fig:H HGR matter comparison}
\end{figure}
where $\rho_{\Lambda}$ is the (constant) vacuum energy density. It is clear from eq.~\eqref{eq:quad pot example}, that $E^{\text{H}}$ and $E^{\phi}$ can only be equivalent (by which we mean, equal up to a multiplicative function of $X$) if \smash{$\rho_{\text{mat}}+P_{\text{mat}}=0$}. This is what we require, that the scalar field only screens the vacuum energy, and not all forms of matter and radiation. Both the Hubble parameter $H$ and the scalar field $\phi$ will respond to $\rho_{\text{mat}}$ and any energy density that does not have a dark energy equation of state, i.e., \smash{$w=\frac{P}{\rho}=-1$}. As the scalar field is not directly coupled to matter in this model, one sees from (\ref{eq:continuity}) that the matter energy density will decay according to \smash{$\rho_{\text{mat}}\propto a^{-3(1+w_{\text{mat}})}$}, allowing for a conventional matter dominated epoch, and furthermore, for eqs.~\eqref{eq:Hubble eom quad pot} and~\eqref{eq:scalar eom quad pot} to be asymptotically equivalent as $\rho_{\text{mat}}\to 0$.

We wish to show now, that this example model possess a viable cosmological history, for which \smash{$H^{2}\propto a^{-3}$} during pressureless matter domination. Note that it is not automatically true that this is possible, even though $H$ responds to a non-zero matter component $\rho_{\text{mat}}$. To determine whether this model does admit a viable cosmological history, we need to consider the eqs.~\eqref{eq:Friedmann constraint quad pot},~\eqref{eq:Hubble eom quad pot} and~\eqref{eq:scalar eom quad pot}, together with the continuity equation (\ref{eq:continuity}).
\begin{figure}[t!]
 \begin{center}
   \includegraphics[scale=0.83]{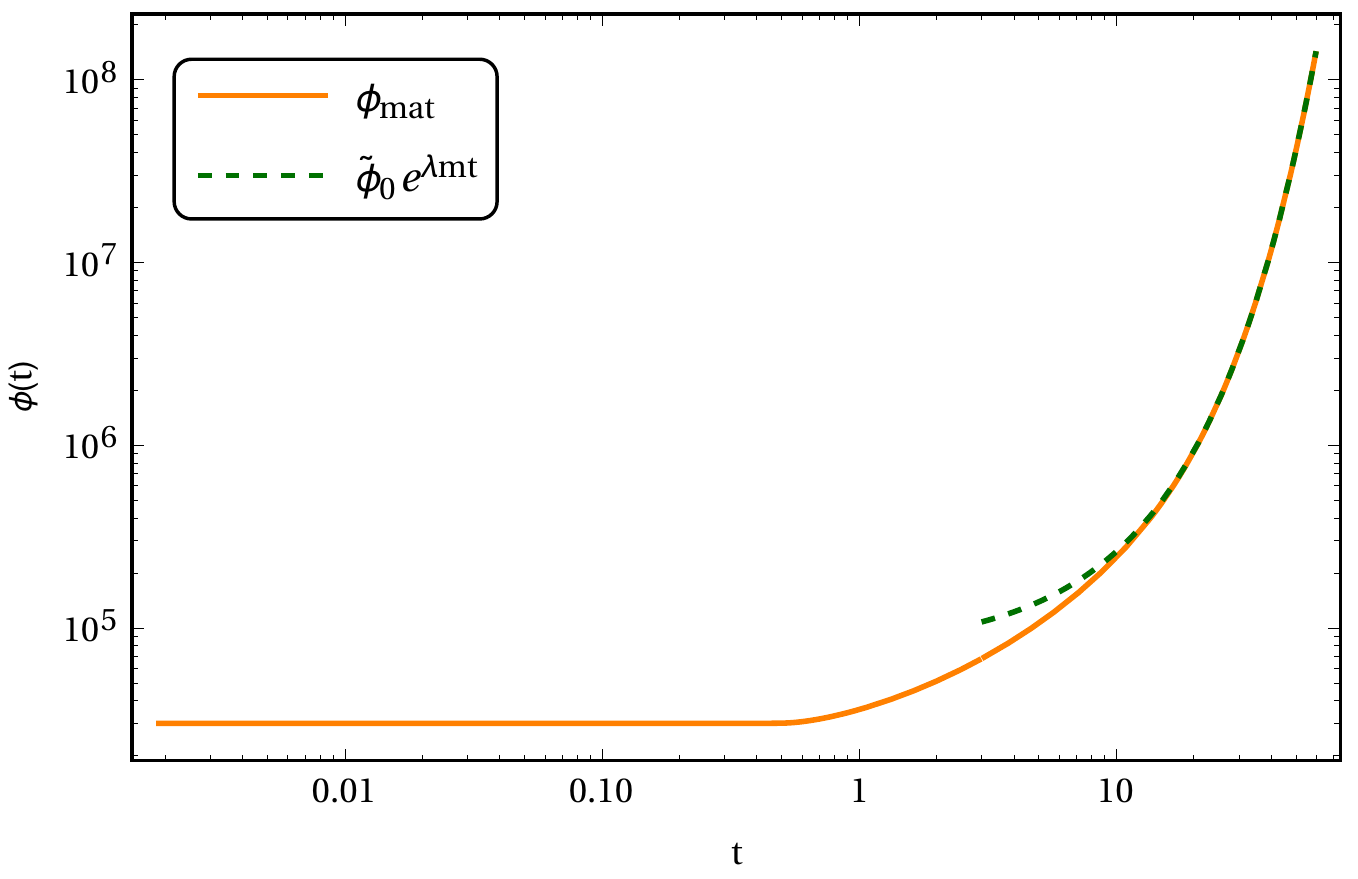}
  \end{center}
  \caption{The corresponding evolution of the scalar field $\phi$ (orange line), numerically evolved using eqs.~\eqref{eq:quad pot example} and~\eqref{eq:continuity}. We see that, as $H$ approaches the de Sitter point (see fig.~\ref{fig:H HGR matter comparison}), $\phi$ transitions to its late time de Sitter solution, which agrees well with the approximate analytic solution $\tilde{\phi}_{0}\:e^{\lambda mt}$ (dashed green line), where $\tilde{\phi}_{0}\simeq 7.4\times 10M_{\text{Pl}}$.}\label{fig:numeric analytic scalar matter case}
\end{figure}

As in the vacuum case, to solve the system of equations [eq.~\eqref{eq:quad pot example}] numerically, we first solve the Friedmann constraint [eq.~\eqref{eq:Friedmann constraint quad pot}] and Hubble equation of motion [eq.~\eqref{eq:Hubble eom quad pot}] for $H$ and $\dot{H}$, respectively. We then insert these expressions back into the scalar equation of motion [eq.~\eqref{eq:scalar eom quad pot}], thus eliminating $H$ and $\dot{H}$ from eq.~\eqref{eq:scalar eom quad pot}. Given this, we then note that we seek solutions in which $H$ closely mimics the standard matter era of GR. To this end, we choose the following initial values for $\rho_{\text{mat}}$, $\phi$ and $\dot{\phi}$: $\rho_{\text{mat},i}=10^{-1}M_{\text{Pl}}^{4}$, $\phi_{i}=30M_{\text{Pl}}$ and $\dot{\phi}_{i}=10^{-9}M_{\text{Pl}}^{2}$. Furthermore, we keep the free parameters of the model set as they were in the vacuum case (cf.~eq.~\eqref{eq:parameter choices}). Figure~\ref{fig:H HGR matter comparison} shows a plot of evolution of the Hubble parameter $H$ of this model (solid orange line), calculated numerically from the full equations of motion, and the GR solution for a matter dominated epoch $H_{\text{GR}}=\frac{2}{3t}$ (dashed green line), as functions of $t$, from $t_{i}\simeq 2\times 10^{-3}M^{-1}$ to $t_{f}= 60M^{-1}$. We see that the full solution closely follows the matter dominated GR solution for $t\ll\frac{1}{m}$, before transitioning to a period of accelerated expansion. Importantly, it is clear, that as $\rho_{\text{mat}}\to 0$ and $\rho_{\Lambda}$ starts to dominate the energy density, $H$ departs from $H_{\text{GR}}$, transitioning to the de Sitter state, where it asymptotes to $H=h$ (dashed black line in Fig.~\ref{fig:H HGR matter comparison}). Note the ``bump'' in the $H$ as it initially departs from the GR solution. This is to be expected from observing the behaviour of $H$ in the vacuum case (cf.~Fig.~\ref{fig:quad pot exp.}). Indeed, for cases in which $H_{i}\sim 10h$ in the vacuum case, this ``bump'' feature appears in the full numerical solution for $H$ (cf.~Fig.~\ref{fig:quad pot exp.}). 

Moreover, we expect that as $H$ transitions to a period of accelerated expansion, the scalar field should join its asymptotic solution $\phi(t)\approx\tilde{\phi}_{0}\:e^{\lambda mt}$. We see from Fig.~\ref{fig:numeric analytic scalar matter case}, that this is indeed the case; the late-time behaviour of the full numerical solution for the scalar field (solid orange line) is in good agreement with the expected behaviour $\phi(t)\approx\tilde{\phi}_{0}\:e^{\lambda mt}$ (dashed green line). Therefore, this example model admits a period of matter domination, in which the dynamics of $H$ is dominated by a dust component $\rho_{\text{mat}}$. Moreover, it is found that this example model can support a radiation dominated epoch (i.e.~$w_{\text{mat}}=\frac{1}{3}$) that follows the GR solution at early times ($t\ll\frac{1}{m}$), before asymptoting to $H=h$. Given that the GR solution for $H$ differs only by a numerical factor relative to the matter dominated case, one expects that the full solution for $H$ should exhibit behaviour very similar to that given in Fig.~\ref{fig:H HGR matter comparison}. Indeed, this is precisely what is found, confirming that this example model can exhibit a standard cosmological evolution at early times. 
\\
\subsection{Stability through a phase transition}
In the previous two subsections, we have been able to show through a combination of analytic and numerical analyses, that this example model admits self-tuning solutions for the scalar field $\phi$, such that in vacuum it can screen the cosmological constant. Furthermore, it can support a period of matter domination in which the Hubble parameter $H$ closely follows the GR solution at early times, before transitioning to a period of late-time accelerated expansion, where $H$ asymptotes to the de Sitter point $H=h$. 

We now wish to examine the behaviour of $H$ in the case where the vacuum energy undergoes a phase transition, $\rho_{\Lambda 1}\to\rho_{\Lambda 2}$. One of the attributes of the tempered self-tuning mechanism is that it only responds to constant energy densities. Since the vacuum energy $\rho_{\Lambda}$ will vary in time during the transitional period (i.e.~$w\neq -1$), it is therefore interesting to study how self-tuning is affected in such a scenario. A priori, one expects that it will briefly pause, such that $\delta H$ leaves its initial late-time trajectory. As $\rho_{\Lambda}$ settles down to its new constant value (such that $w\to -1$), then self-tuning should recommence and $\delta H$ should transition to a new late-time attractor trajectory, such that $\delta H\to 0$ asymptotically. 

Since, in this example model, it is not possible to start out on the de Sitter solution, it will generally be the case that the phase transition occurs whilst $H$ is tending towards the attractor solution $H=h$. To carry out a numerical analysis, therefore, we shall return to the original equations for this example model [eq.~\eqref{eq:quad pot example}], this time including the continuity equation for (the now time-dependent) $\rho_{\Lambda}$:
\begin{equation}\label{eq:continuity eq phase}
 \dot{\rho}_{\Lambda} \:+\: 3H\big[\rho_{\Lambda} \:+\: P_{\Lambda}\big]=0\;.
\end{equation}
Following in the steps of the original analysis conducted by Appleby and Linder~\cite{Appleby:2018yci}, we model the transition in $\rho_{\Lambda}$ numerically as
\begin{equation}
 \rho_{\Lambda} \ = \ \rho_{\Lambda 1} \:-\: \frac{(\rho_{\Lambda 1} \:-\: \rho_{\Lambda 2})}{2}\bigg[1 \:+\: \text{tanh}\Big(M\frac{t-t_{p}}{\epsilon}\Big)\bigg]\;,
\end{equation}
with $\rho_{\Lambda 1}=10^{-4}M_{\text{Pl}}^{2}$, $\rho_{\Lambda 2}=10^{-5}M_{\text{Pl}}^{2}$, $M=10^{-3}M_{\text{Pl}}$, $t_{p}=30M^{-1}$ and $\epsilon=0.2$. The effective pressure $P_{\Lambda}$ is then given by,
\begin{equation}
 P_{\Lambda} \ = \ -\: \rho_{\Lambda} \:-\: \frac{\dot{\rho}_{\Lambda}}{3H}\;.
\end{equation}
This choice for $\rho_{\Lambda}$ allows for it to smoothly evolve from an initial value $\rho_{\Lambda}\simeq\rho_{\Lambda 1}$ to a final value $\rho_{\Lambda}\simeq\rho_{\Lambda 2}$ for $t>t_{p}$. 

\begin{figure}[t!]
 \begin{center}
   \includegraphics[scale=0.85]{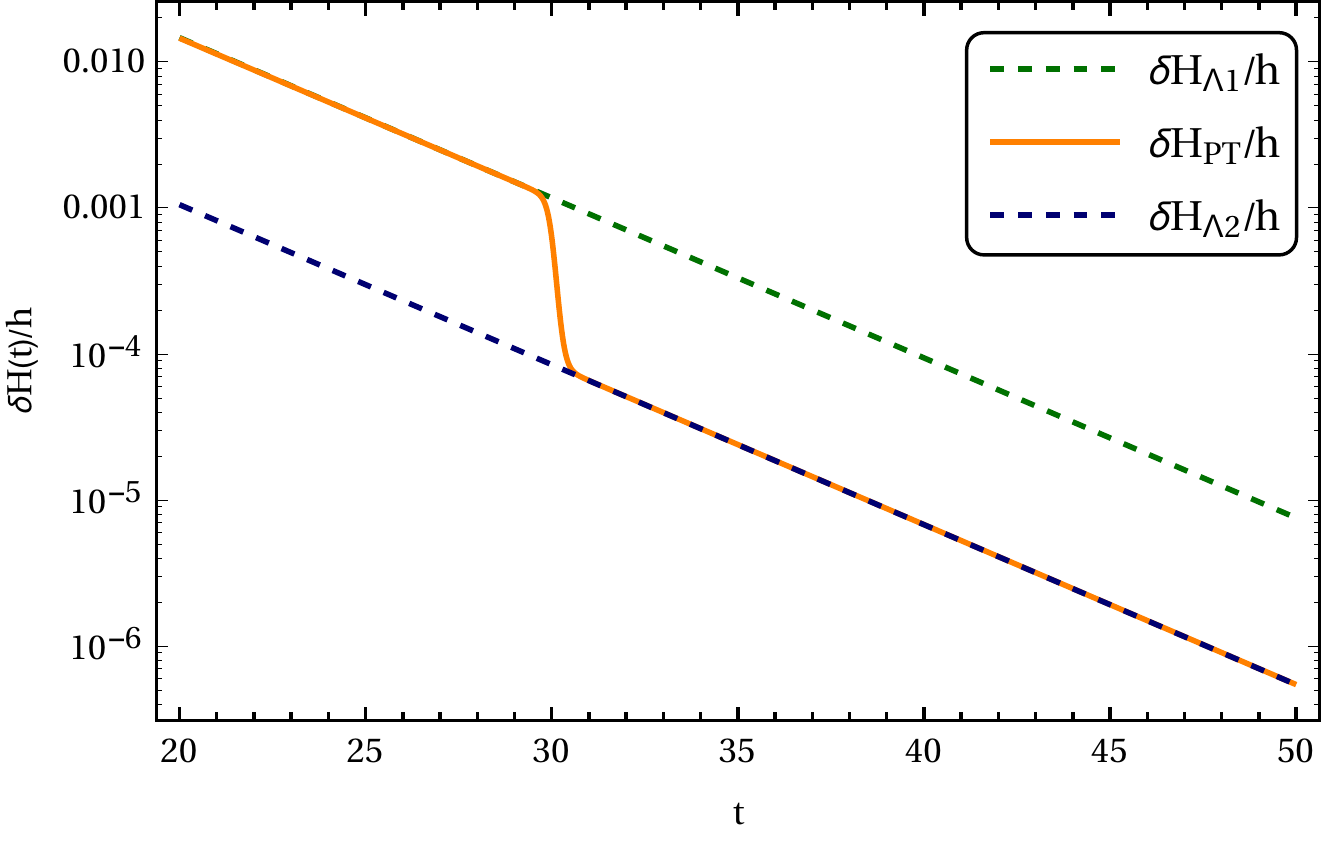} 
  \end{center}
  \caption{Late-time evolution of $\frac{\delta H_{\text{PT}}}{h}$ through a phase transition of the vacuum energy (solid orange line). The dashed green and blue lines show the late-time analytic evolution of $\frac{\delta H}{h}$ before and after the phase transition, respectively. We see that $\frac{\delta H_{\text{PT}}}{h}$ smoothly evolves between the two (green to blue), such that  it tracks the dashed blue line as it tends towards its asymptotic limit $\frac{\delta H}{h}\to 0$.}\label{fig:Hubble phase transition}
\end{figure}
As remarked upon in the previous subsection, since we cannot start out on the de Sitter solution, we shall choose initially conditions such that $H(t_{i})=h+\delta H(t_{i})$. We have chosen the time $t_{p}$ such that the phase transition occurs at a suitably late point and $H$ is already asympotically approaching its de Sitter solution. This will allow us to assess the stability of the self-tuning limit under a phase transition. As before, we shall keep the remaining free parameters fixed as $h=10^{-3}M_{\text{Pl}}$ and $m=10^{-4}M_{\text{Pl}}$. Solving the Friedmann constraint [eq.~\eqref{eq:Friedmann constraint quad pot}] and Hubble equation of motion [eq.~\eqref{eq:Hubble eom quad pot}] for $H$ and $\dot{H}$, respectively, we then insert these solutions into the scalar field equation [eq.~\eqref{eq:scalar eom quad pot}]. We then numerically evolve the resulting equation of motion, taking the initial values for $\phi$ and $\dot{\phi}$ to be $(\phi_{i},\,\dot{\phi}_{i})=(10M_{\text{Pl}},\, 10^{-2}M_{\text{Pl}}^{2})$. In Fig.~\ref{fig:Hubble phase transition}, we show a plot of the late-time behaviour of the numerical solution for $\frac{\delta H}{h}$ obtained from the full equations of motion (solid orange line). Moreover, using the analytic expression $\frac{\delta H}{h}$ with the late-time analytic solution for $\phi$ (cf.~eqs.~\eqref{eq:approx phi} and~\eqref{eq:deltaH sol2}) inserted, we plot the associated trajectories for $\rho_{\Lambda}=\rho_{\Lambda 1}$ (dashed green line) and $\rho_{\Lambda}=\rho_{\Lambda 2}$ (dashed blue line). Observe that $\frac{\delta H}{h}$ intially starts out on the analytic trajectory corresponding to $\rho_{\Lambda}=\rho_{\Lambda 1}$. At the phase transition $t_{p}=30M^{-1}$, we see that it departs from this trajectory, smoothly transitioning to the analytic trajectory corresponding to $\rho_{\Lambda}=\rho_{\Lambda 2}$, upon which it asymptotes to the de Sitter solution $\frac{\delta H}{h}\to 0$. Throughout this late-time evolution, the scalar field follows an exponentially growing trajectory (cf.~Fig.~\ref{fig:numeric analytic scalar}) which agrees with the late-time analytic solution [eq.~\eqref{eq:approx phi}].

\section{Conclusions}
\label{sec:conclusions}
In this paper, we have presented an extension of the well-tempered proposal for dealing with the cosmological constant problem \cite{Appleby:2011aa} by providing a very general solution to the constraints imposed by the well-tempered mechanism. We started within the framework of Horndeski's scalar-tensor theory \cite{Horndeski:1974wa}, ensuring that the current gravitational wave constraints were observed. We then required that the scalar had a canonical kinetic term with a general potential, while also allowing for a general $G_4(\phi)$, and found that well-tempering is possible if $G_3(\phi,X)$ takes the explicit form given in eq.~\eqref{eq:G3 sol}. In order to see this general solution work in practise we presented a series of numerical experiments, showing that the vacuum energy is indeed hidden from the spacetime geometry and, importantly, that we are able to construct a sensible cosmological history with a fluid dominated period before the de Sitter phase.

There are a number of routes that should now be followed in order to explore the solution in this paper. Given that we have a number of arbitrary functions in our model, it would be useful if constraints could be placed on them. For example, one should examine how scalar and tensor perturbations behave for different models, in order to rule out unstable cases. Another important avenue of research is to calculate quantum corrections, to find out how robust the structure is when radiative corrections are included. Also, given that we have a scalar field in the picture, we need to make sure that any new force due to the scalar is within the confinement of the Solar System and laboratory constraints of gravity. 

~\\ 
\noindent{\bf Acknowledgements}
WTE and PMS are supported by STFC Grant No. ST/L000393/1 and ST/P000703/1. SYZ acknowledges support from the starting grant from University of Science and Technology of China (Grant No.: KY2030000089) and the National 1000 Young Talents Program of China (Grant No.: GG2030040375).	
\\
\begin{appendices}
 \section{Deriving the well-tempered form of $G_{3}(\phi,X)$}\label{sec:G3 calc}
	Here we present a more detailed discussion of how to determine the coefficient functions $f_{m}$ of the Taylor series representation of $G_{3}$. Furthermore, we shall show that the resulting expression for $G_{3}$ can be recast into the form given in eq.~\eqref{eq:G3 sol}. With this in mind, let us consider the case where we truncate $G_{4}$ and $V$ to quadratic order (i.e., $n=2$ and $l=2$), and $G_{3}$ to quartic order (i.e., $m=4$). This leaves us with a tower of coupled differential equations for $f_{4}$, $f_{3}$, $f_{2}$, $f_{1}$ and $f_{0}$. There is a single equation which is de-coupled from the rest, dependent solely on $f_{4}(X)$ and its derivatives:
	\begin{equation}\label{eq:f4 diff eq}
	2Xf''_{4} \: + \: 3f'_{4} \ = \ 0 \;.
	\end{equation}
	It is therefore most straightforward to start by solving this equation and then the remaining in succession, given this solution. The general solution to eq.~\eqref{eq:f4 diff eq} is
	\begin{equation}\label{eq:f4 sol}
	f_{4}(X) \ = \ -\frac{2\tilde{\kappa}_{1}}{\sqrt{2X}} \: + \: \kappa_{1}\;,
	\end{equation}
	where $\tilde{\kappa}_{1}$ and $\kappa_{1}$ are arbitrary integration constants. We choose our boundary conditions such that $\tilde{\kappa}_{1}=0$, but keep $\kappa_{1}$ arbitrary for now.\footnote{Note that we could have chosen $\tilde{\kappa}_{1}\neq 0$, however, this leads to a significantly more complicated set of solutions for the remaining $f_{m}$'s, such that $G_{3}$ cannot be cast into the elegant form given by eq.~\eqref{eq:G3 sol}.} Given eq.~\eqref{eq:f4 sol}, we can then solve the differential equation for $f_{3}$:
	\begin{flalign}\label{eq:f3 sol}
	&6hX\big[3hXf'_{3} \: - \: 4\kappa_{1}\sqrt{2X}\big]f''_{3} \: + \: 27h^{2}X(f'_{3})^{2} \: - \: 48h\kappa_{1}\sqrt{2X}f'_{3} \: + \: 32\kappa_{1}^{2} \ = \ 0 \nonumber\\[0.8em] &\Rightarrow\qquad f_{3}(X) \ = \kappa_{2} \: + \: \frac{4\kappa_{1}}{3h}\sqrt{2X} \;,
	\end{flalign}
	where $\kappa_{2}$ is an arbirary integration constant. Having determined expressions for $f_{4}$ and $f_{3}$, we are led to the following differential equation for $f_{2}$:
	\begin{equation}
	h(2X)^{3/2}f''_{2} \: + \: 3h\sqrt{2X}f'_{2} \: - \: \frac{4\kappa_{1}}{h}\sqrt{2X} \: - \: 2\kappa_{2} \ = \ 0 \:,
	\end{equation}
	the solution to which is 
	\begin{equation}\label{eq:f2 sol}
	f_{2}(X) \ = \frac{4\kappa_{1}}{3h^{2}}X \: + \: \frac{\kappa_{2}}{h}\sqrt{2X} \: - \: \frac{2\kappa_{3}}{\sqrt{2X}} \: + \: \kappa_{4} \;, 
	\end{equation}
	where $\kappa_{3}$ and $\kappa_{4}$ are again arbitrary integration constants. Then, to determine a solution for $f_{1}$, we need to solve:
	\begin{flalign}
	 0 \ =& \ 18\sqrt{2}h^{3}X^{2}f''_{1} \: + \: 27\sqrt{2}h^{3}Xf'_{1} \: - \: 32\kappa_{1}X^{3/2} \: - \: 18\sqrt{2}h\kappa_{2}X \: - \: 12h^{2}\kappa_{4}\sqrt{X} \nonumber\\[0.3em] & \ - \: 12\sqrt{2}h^{3}c_{2} \: - \: \sqrt{2}hV_{2} \;.
	\end{flalign}
	This yields the solution
	\begin{equation}\label{eq:f1 sol}
	f_{1}(X) \ = \ \kappa_{6} \: - \: \frac{2\kappa_{5}}{\sqrt{X}} \: + \: \frac{2\kappa_{4}}{3h}\sqrt{2X} \: + \: \frac{2\kappa_{2}X}{3h^{2}} \: + \: \frac{4\kappa_{1}}{27h^{3}}(2X)^{3/2} \: + \: \frac{(12h^{2}c_{2} \: + \: V_{2})}{9h^{2}}\,\text{log}(X) \;,
	\end{equation}
	with $\kappa_{5}$ and $\kappa_{6}$ arbitrary integration constants. Finally, we turn our attention to the remaining differential equation for $f_{0}(X)$:
	\begin{flalign}\label{eq:f0 diff eq}
	0 \ =& \ 12h\kappa_{1}(2X)^{5/2}f''_{0} \: + \: 36h\kappa_{1}(2X)^{3/2}f'_{0} \: - \: \frac{64\kappa_{1}}{3}\Big[c_{2} \: + \: \frac{V_{2}}{12h^{2}}\Big]X\,\text{log}(X) \: - \: \frac{80\kappa_{1}^{2}}{27h^{3}}(2X)^{5/2} \nonumber\\[0.8em] & - \: \frac{64\kappa_{1}\kappa_{2}}{3h^{2}}X^{2} \: - \: \frac{8\kappa_{1}\kappa_{4}}{h}(2X)^{3/2} \: - \: \frac{8\kappa_{1}}{9h^{2}}\big[4V_{2} \: + \: 12\sqrt{2}h\kappa_{3} \: + \: 3h^{2}(6\kappa_{6} \: + \: 4c_{2} \: - \: 3) \big]X \nonumber\\[0.8em] & - \: \frac{4\kappa_{1}}{3h}\big[V_{1} \: + \: 12h^{2}c_{1}\big]\sqrt{2X} \: - \: 8\kappa_{3}\big[V_{2} \: + \: 6\sqrt{2}h\kappa_{3} \: - \: 6h^{2}c_{2}\big]\frac{1}{\sqrt{X}}	\end{flalign}
	Although looking rather unpleasant, eq.~\eqref{eq:f0 diff eq} does have a reasonable solution:
	\begin{flalign}\label{eq:f0 sol}
	f_{0}(X) \ =& \ \kappa_{8} \: + \: \frac{\kappa_{3}}{6h\kappa_{1}}\big[12h\kappa_{3} \: + \: \sqrt{2}V_{2} \: - \: 6\sqrt{2}h^{2}c_{2}\big]\frac{1}{X} \: - \: \frac{2\kappa_{7}}{\sqrt{X}} \nonumber\\[0.3em] &  + \: \big[24h\kappa_{3} \: - \: 6\sqrt{2}V_{2} \: - \: -3\sqrt{2}h^{2}(3 \: + \: 20c_{2} \: - \: 6\kappa_{6})\big]\frac{\sqrt{X}}{54h^{3}} \: + \: \frac{2\kappa_{4}}{9h^{2}}X \: + \: \frac{\kappa_{2}}{27h^{3}}(2X)^{3/2} \nonumber\\[0.3em] & + \: \frac{\kappa_{1}}{81h^{4}}(2X)^{2} \: + \: \frac{2}{3}\Big[c_{1} \: + \: \frac{V_{1}}{12h^{2}}\Big]\,\text{log}(X) \: + \: \frac{4}{9h}\Big[c_{2} \: + \: \frac{V_{2}}{12h^{2}}\Big]\sqrt{2X}\,\text{log}(X) \;,
	\end{flalign}
	with $\kappa_{7}$ and $\kappa_{8}$ arbitrary integration constants.
	
	Given eqs.~\eqref{eq:f4 sol},~\eqref{eq:f3 sol},~\eqref{eq:f2 sol},~\eqref{eq:f1 sol} and~\eqref{eq:f0 sol}, we can reconstruct $G_{3}$ by inserting their expressions into eq.~\eqref{eq:G3 expansion}. Note, however, that not all of the coefficients $\kappa_{i}$ ($i=1,\ldots ,8$) can be arbitrary. Indeed, one must fix two of them to be
	\begin{subequations}
		\begin{flalign}
		\kappa_{3} \ =& \ \frac{6h^{2}c_{2} \: - \: V_{2}}{6\sqrt{2}h} \;, \\[0.8em] \kappa_{5} \ =& \ \frac{6h^{2}c_{1} \: - \: V_{1}}{6\sqrt{2}h} \;,
		\end{flalign}
	\end{subequations}
	such that $f_{4}$, $f_{3}$, $f_{2}$, $f_{1}$ and $f_{0}$ reduce to
	\begin{subequations}
		\begin{flalign}
		f_{4}(X) \ =& \ \kappa_{1} \;,\\[0.8em] f_{3}(X) \ =& \ \kappa_{2} \: + \:  \frac{\kappa_{1}}{3h}\sqrt{2X} \;,\\[0.8em] f_{2}(X) \ =& \ \kappa_{4} \: - \: \frac{6h^{2}c_{2} \: - \: V_{2}}{3h\sqrt{2X}} \: + \: \frac{\kappa_{2}}{h}\sqrt{2X} \: + \: \frac{4\kappa_{1}}{3h^{2}}X \;,\\[0.8em] f_{1}(X) \ =& \ \kappa_{6} \: - \: \frac{6h^{2}c_{2} \: - \: V_{2}}{3h\sqrt{2X}} \: + \: \frac{2\kappa_{4}}{3h}\sqrt{2X} \: + \: \frac{2\kappa_{2}}{3h^{2}}X\: + \: \frac{4\kappa_{1}}{27h^{3}}(2X)^{3/2} \: + \: \frac{4}{3}\Big[c_{2} \: + \: \frac{V_{2}}{12h^{2}}\Big]\,\text{log}(X) \;, \\[0.1em] f_{0}(X) \ =& \ \kappa_{8} \: - \: \frac{2\kappa_{7}}{\sqrt{X}} \: - \: \big[4V_{2} \: + \: 3h^{2}(3 \: + \: 16c_{2} \: - \: 6\kappa_{6})\big]\frac{\sqrt{2X}}{54h^{3}} \:  + \: \frac{2\kappa_{4}}{9h^{2}}X \: + \: \frac{\kappa_{2}}{27h^{3}}(2X)^{3/2} \nonumber\\[0.5em] & + \: \frac{\kappa_{1}}{81h^{4}}(2X)^{2} \: + \: \frac{2}{3}\Big[c_{1} \: + \: \frac{V_{1}}{12h^{2}}\Big]\,\text{log}(X) \: + \: \frac{4}{9h}\Big[c_{2} \: + \: \frac{V_{2}}{12h^{2}}\Big]\sqrt{2X}\,\text{log}(X) \;.
		\end{flalign}
	\end{subequations}
	Inserting these expressions back into the series expansion for $G_{3}$, we find
	\begin{flalign}\label{G3 truncated sol}
	G_{3}(\phi,X) \ =& \ -2h\Big[\big(c_{1}\phi \: + \: c_{2}\phi^{2}\big) \: + \: \frac{1}{6h^{2}}\big(X \: - \: V_{1}\phi \: - \: V_{2}\phi^{2}\big) \Big]\frac{1}{\sqrt{2X}} \: - \: \frac{2\kappa_{7}}{\sqrt{X}}  \nonumber\\[0.3em] & \ - \: \frac{8}{9h}\Big[c_{2} \: + \: \frac{V_{2}}{12h^{2}}\Big]\sqrt{2X} \: + \: \frac{2}{3}\Big[\big(c_{1} \: + \: 2c_{2}\phi\big) \: + \: \frac{1}{12h^{2}}\big(V_{1} \: + \: 2V_{2}\phi\big)\Big]\,\text{log}(X) \nonumber\\[0.3em] & \ + \:  \frac{4}{9h}\Big[c_{2} \: + \: \frac{V_{2}}{12h^{2}}\Big]\sqrt{2X}\,\text{log}(X) \: + \: \Big\lbrace\kappa_{8} \: + \: \kappa_{6}\Big[\phi \: + \: \frac{\sqrt{2X}}{3h}\Big] \: + \: \kappa_{4}\Big[\phi \: + \: \frac{\sqrt{2X}}{3h}\Big]^{2} \nonumber\\[0.3em] & \ + \: \kappa_{2}\Big[\phi \: + \: \frac{\sqrt{2X}}{3h}\Big]^{3} \: + \: \kappa_{1}\Big[\phi \: + \: \frac{\sqrt{2X}}{3h}\Big]^{4}\Big\rbrace \;.
	\end{flalign}
	Notice that the terms in the braces are clearly the quartic order truncation of an arbitrary function of \smash{$u=\phi +\frac{\sqrt{2X}}{3h}$}. Further identifying the remaining terms in eq.~\eqref{G3 truncated sol} as truncations of combinations of $K$, $V$, and $G_{4}$, it is found that $G_{3}$ can be recast into the following elegant form:
	\begin{flalign}\label{eq:G3 well-tempered}
	G_{3}(\phi,X) \ =& \ - \: \frac{X \: - \: V(\phi)}{3h\sqrt{2X}} \: - \: \frac{2hG_{4}(\phi)}{\sqrt{2X}} \: + \: \frac{2}{3}\frac{\partial}{\partial\phi}\Big[G_{4}(u) \: + \: \frac{1}{12h^{2}}V(u)\Big]\,\text{log}\bigg(\frac{X}{\mu^{4}}\bigg)  \nonumber\\[0.8em] &  + \: \Big[\frac{a_{0}h}{\sqrt{2X}} \: + \: \mathcal{F}(u)\Big] \: - \: \frac{4}{3}\int_{0}^{1}\frac{\frac{\partial}{\partial\phi}\Big[\Big(G_{4}(u) \: + \: \frac{1}{12h^{2}}V(u)\Big) \: - \: \Big(G_{4}(\tilde{u}) \: + \: \frac{1}{12h^{2}}V(\tilde{u})\Big)\Big]}{1-s}\,\sd{d}s \;,
	\end{flalign}
	where $\mathcal{F}(u)$ is an arbitrary function, with the variables $u$ and $\tilde{u}$ given by eq.~\eqref{eq:variables def}, and we have introduced an arbitrary mass scale $\mu$ such that the logarithm is dimensionally correct (the additional contribution introduced in doing so can be absorbed into the arbitrary function $\mathcal{F}(u)$). Note further, that we have absorbed $c_{0}$ and $V_{0}$ into the arbitrary coefficient $\kappa_{7}$, redefining it such that $\kappa_{7}\to a_{0}h\equiv 2hc_{0} - \frac{V_{0}}{3h} - 2\sqrt{2}\kappa_{7}$.
	
	One can then repeat this analysis for higher order truncations of the Taylor series in eq.~\eqref{eq:G4 G3 expansion},\footnote{We went up to order twenty in our calculations.} and in each case one can resum these $f_{m}(X)$, such that we obtain an expression for $G_{3}$ given by eq.~\eqref{eq:G3 sol}. Importantly, it can be shown (by inserting eq.~\eqref{eq:G3 well-tempered} into eq.~\eqref{eq: de Sitter constraint}) that the fully resummed expression for $G_{3}$ satisfies the well-tempered constraint [eq.~\eqref{eq: de Sitter constraint}], thus providing confirmation that eq.~\eqref{eq:G3 well-tempered} is the correct resummation of its corresponding Taylor expansion [eq.~\eqref{eq:G3 expansion}]. 
\end{appendices}

\end{document}